\documentclass[osajnl,twocolumn,showpacs,superscriptaddress,10pt]{revtex4-1} %% use 11pt for Applied Optics
\usepackage{amsmath,amssymb,graphicx}
\begin{document}

\title{Spectral broadening and shaping of nanosecond pulses: towards shaping of single photons from quantum emitters}

\author{Imad Agha}\email{Corresponding author: iagha1@udayton.edu}
\affiliation{Center for Nanoscale Science and Technology, National
Institute of Standards and Technology, Gaithersburg, MD 20899}
\affiliation{Maryland NanoCenter, University of Maryland, College Park, MD
20742}
\affiliation{Department of Physics, and Electro-Optics Graduate Program, University of Dayton, Dayton, Ohio 45469}

\author{Serkan Ates}
\affiliation{Center for Nanoscale Science and Technology, National
Institute of Standards and Technology, Gaithersburg, MD 20899}
\affiliation{Maryland NanoCenter, University of Maryland, College Park, MD
20742}

\author{Luca Sapienza}
\affiliation{Center for Nanoscale Science and Technology, National
Institute of Standards and Technology, Gaithersburg, MD 20899}
\affiliation{Maryland NanoCenter, University of Maryland, College Park, MD
20742}

\author{Kartik Srinivasan}\email{kartik.srinivasan@nist.gov}
\affiliation{Center for Nanoscale Science and Technology, National
Institute of Standards and Technology, Gaithersburg, MD 20899}

\begin{abstract} We experimentally demonstrate spectral broadening and
shaping of exponentially-decaying nanosecond pulses via nonlinear
mixing with a phase-modulated pump in a periodically-poled lithium niobate (PPLN)
waveguide. A strong, 1550~nm pulse is imprinted with a temporal phase and
used to upconvert a weak 980 nm pulse to 600 nm while simultaneously
broadening the spectrum to that of a Lorentzian pulse up to 10 times shorter.
While the current experimental demonstration is
for spectral shaping, we also provide a numerical study showing the
feasibility of subsequent spectral phase correction to achieve temporal
compression and re-shaping of a 1~ns mono-exponentially decaying pulse to a 250 ps
Lorentzian, which would constitute a complete spectro-temporal waveform
shaping protocol. This method, which uses quantum frequency
conversion in PPLN with $>100:1$ signal-to-noise ratio, is compatible with single photon states of light.

\end{abstract}

%\OCIS{(350.4238); (270.0270); (130.7405).}

\maketitle %% required

\section{Introduction}

Hybrid quantum networks that combine different physical systems
are one approach to achieving the varied functions needed in photonic
quantum information processing systems~\cite{ref:Kimble_Nat08}. Unfortunately, the
disparate components of such a quantum network may not
share the same spectro-temporal properties, leading to an intrinsic
incompatibility that can only be overcome via an "adaptation interface."  For
example, single photon sources based on single quantum emitters like
InAs quantum dots~\cite{ref:Michler_book_2009}, nitrogen vacancy centers in
diamond~\cite{ref:Kurtsiefer}, and neutral alkali atoms~\cite{ref:McKeever}
exhibit desirable features such as on-demand generation with the potential
for high single photon purity. To interface the emission wavelengths below
1000~nm with the low-loss telecommunications band, quantum frequency conversion
interfaces~\cite{ref:Kumar_OL,ref:Raymer_Srinivasan_PT} have been proposed
and developed, both in bulk and on-chip geometries.  However,
wavelength incompatibility is not the only challenge that needs to be overcome. While the temporal waveform of such
two-level quantum emitters is typically a few nanosecond mono-exponential
decay, telecommunications networks are better suited to Gaussian and
square pulses that are much shorter in duration.  Moreover, the bandwidth of
quantum memories that are an integral part of the quantum repeater
protocol may be either broader or narrower than the bandwidth of the photons~\cite{ref:Reim_NatPhot10}.
Spectro-temporal shaping of single photons, while preserving their quantum nature,
is thus a vital tool for hybrid quantum networks~\cite{ref:Silberhorne_pulse_shaper,ref:Raymer_temporal_selection}.

Here, we simultaneously frequency translate and convert the
spectrum of a mono-exponentially decaying pulse to that of a shorter Lorentzian
 wavepacket.  The core of this approach is frequency conversion using a spectrally chirped pump in
a nonlinear medium~\cite{ref:Kielpinski_PRL_2011}.  We focus on
classical input pulses with a mono-exponential decay in the nanosecond
regime, mimicking the emission of the aforementioned single quantum emitters.
We demonstrate spectral broadening of such waveforms using a variation of a setup previously used
in nearly background-free quantum frequency conversion of single
photon from a semiconductor quantum dot~\cite{ref:Ates_Srinivasan_PRL}.  Our
results complement recent work on spectral compression of single photons
through nonlinear wave mixing~\cite{ref:Lavoie}. Moreover, they constitute the
first step towards complete spectro-temporal waveform shaping as proposed by Kielpinski
and colleagues~\cite{ref:Kielpinski_PRL_2011}.  We highlight the experimental requirements
for this, by calculating the spectral phase correction
that would need to be applied to our output field to achieve temporal pulse compression.

\vspace{-0.05in}

\section{Basic principles}

There are several contexts in which spectro-temporal control of
single photon wavepackets has been demonstrated.  Three-level atoms
enable waveform shaping during single
photon generation~\cite{ref:Keller}, while photon echo quantum memories have
been used for pulse compression and
decompression~\cite{ref:Moiseev_Tittel_Qmem}.  Photons with sufficiently
broad spectral bandwidths are amenable to pulse shaping techniques based
on dispersive optics and spatial light modulators, as has
been demonstrated for ultrafast biphoton
wavepackets~\cite{ref:Silberberg_PRL,ref:Lukens_Weiner}. Unfortunately, for
single photon emitters such as quantum dots, trapped atoms, or nitrogen vacancy
centers, the emission bandwidth is too narrow ($ < 1$ GHz) for typical
line-by-line shaping. For such systems, spectral broadening has to precede temporal waveform shaping.

For arbitrary waveform shaping, controlling both the temporal phase (to change the
spectral shape) and the spectral phase (to control the temporal
waveform) is needed.  An approach to doing so, outlined in
Ref.~\onlinecite{ref:Kielpinski_PRL_2011}, starts by imparting a
temporal phase on an incoming single photon wavepacket
to convert the input spectrum to the desired output spectrum. This is
followed by spectral phase compensation to remove the unwanted
accumulated spectral phase, and convert the temporal waveform
to the desired output temporal shape.

\begin{figure}[t]
\centerline{\includegraphics[width=\linewidth]{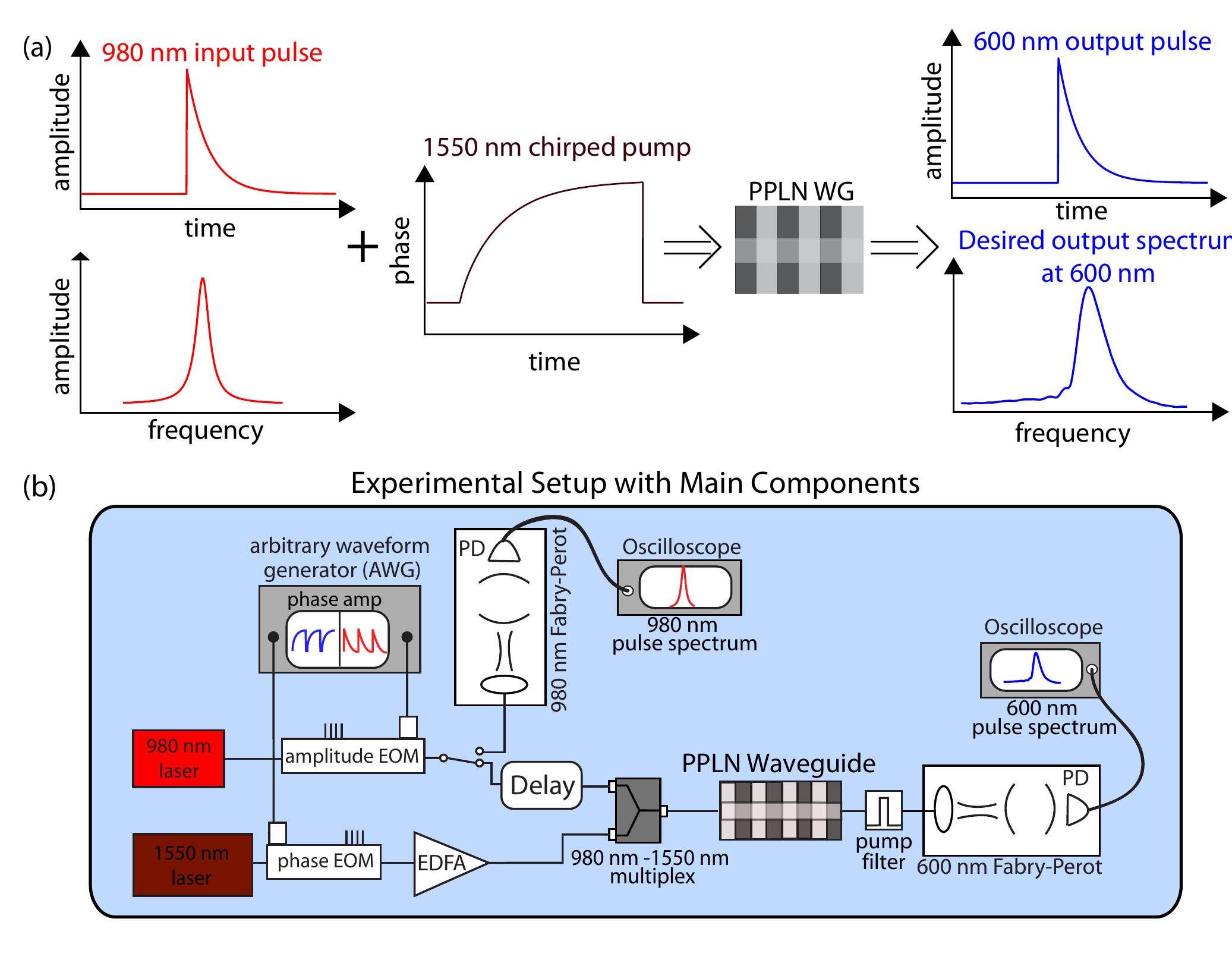}} \vspace{-0.1in}\caption{(a)
Spectral shaping approach: a mono-exponentially
decaying pulse with a Lorentzian spectrum is mixed in a PPLN waveguide with a pump to which a specified
temporal phase is applied. The output light has the same time-domain amplitude, but has been shaped
to match a desired spectrum. (b) Setup: a 980 nm laser passes through an intensity electro-optic modulator
(EOM) controlled by the first channel of an arbitrary waveform generator (AWG), and its spectrum is measured with a Fabry-Perot analyzer (FP).
At the same time, a 1550 nm laser is sent through a phase EOM (to add the correct temporal phase) controlled by the second channel of the AWG,
and is amplified in an erbium-doped fiber amplifier (EDFA). The 980 nm pulse is combined with the strong phase-modulated pump
and sent to the PPLN waveguide. The PPLN output is filtered and the spectrally shaped 600~nm wavelength is measured on a second FP analyzer.}\label{fig:Fig1}
\vspace{-0.2in}
\end{figure}

The temporal phase manipulation step, shown schematically in
Fig.~\ref{fig:Fig1}(a), is the focus of this paper. Here, an incoming single
photon pulse is mixed with a strong, phase-modulated classical pump pulse
in a nonlinear crystal. The frequency of the pump is chosen to
convert the photon to the desired output wavelength, while the temporal phase
is imprinted across the frequency converted single photon.  By correctly
choosing the temporal phase, it is possible to convert the input spectrum
into any desired output spectrum through the properties of Fourier
transforms:

\begin{equation}
\big| \Omega_{out}(\omega)\big|=\big| \int dt\Omega_{in}(t)e^{i\phi(t)}e^{i\omega t}\big|
\end{equation}

\noindent When the output spectrum is broader than the input spectrum
(temporal compression), the method of stationary phase can produce an analytical solution for the temporal phase $\phi(t)$.  For converting a
mono-exponentially decaying input $\Omega_{in}(t)=H(t)e^{-|t|/\tau}$, where $H(t)$ is the Heaviside
step function, to a Gaussian $\Omega_{out}(t)=e^{-t^2/2\sigma^2}$, we have
$\phi(t)\approx\frac{\sqrt{2}}{\sigma}\int_0^{t}{erf}^{-1}(-t^{\prime}/\tau)dt^{\prime}$, where $erf^{-1}(t)$
is the inverse error function~\cite{ref:Kielpinski_PRL_2011}.  While Ref.~\onlinecite{ref:Kielpinski_PRL_2011}
considers large temporal compression ratios ($\approx$100), we focus on more modest ratios we can
achieve experimentally ($\lesssim10$).  Here, this $\phi(t)$ produces a spectrum more closely
matching a temporal Lorentzian, resulting from using the method of stationary phase outside of its strict domain
of validity.

The addition of a temporal phase cannot modify the temporal profile, so that even though
the output waveform has the desired spectrum, its
time-domain waveform is unchanged due to an extra spectral phase imprinted
on the output pulse. This unwanted spectral phase can be extracted numerically (by comparing the spectral output
of the nonlinear mixing stage to that of a transform-limited pulse). If the
broadening in the first step increases the bandwidth sufficiently, standard
spatial light modulator pulse shaping techniques (as in
Refs.~\onlinecite{ref:Silberberg_PRL,ref:Lukens_Weiner}) may be used for
spectral phase compensation, so that at the output, the
pulse should have both the desired temporal and spectral profiles.

\section{Experiment}

In the experiment, we demonstrate spectral waveform shaping of classical
light pulses compatible with the single photon regime. We target
a mono-exponentially decaying pulse in time (a Lorentzian spectrum) of various durations
(1~ns to 30~ns, spanning spontaneous emission lifetimes ranging from InAs/GaAs quantum
dots to trapped atoms, respectively) and spectrally shape it to match the
spectrum of a compressed Lorentzian wavepacket in time (mono-exponential decay in frequency),
while simultaneously shifting the wavelength. The Lorentzian shape has been chosen because the required phase profile
can be implemented with the available resources, where the main limitation is
the 4~GHz sampling rate of our arbitrary-waveform generator and resulting number of phase points
across the pulses.

To start, an exponentially decaying pulse that mimics the temporal
profile of a single quantum emitter is generated by passing an
attenuated 980~nm laser through an intensity electro-optic modulator (EOM) driven by an
arbitrary waveform generator (AWG) (Fig. 1(b)). Simultaneously, the second channel of the AWG drives
a phase EOM and imprints the desired phase on the 1550 nm pump.   The
pump is subsequently amplified via an erbium-doped fiber amplifier and
combined with the 980~nm pulses in a PPLN waveguide.  Sum frequency
generation yields 600~nm light whose spectrum is
broadened and shaped into the spectrum of a Lorentzian pulse in time.
The wall-to-wall conversion efficiency (which includes input and output coupling losses) is estimated
to be $40~\%\pm1~\%$~\cite{ref:spectral_broadening_note} with a signal-to-noise ratio $>100:1$,
as described in Ref.~\onlinecite{ref:Ates_Srinivasan_PRL}, where quantum frequency conversion
with a true single photon source was performed, establishing
that the measurements shown below can be directly extended to the single
photon regime.

The initial 980~nm spectrum is measured by a Fabry-Perot (FP) analyzer
with 10~GHz free spectral range and finesse $>$ 100. The 980 nm~pulse is then
switched into a delay line, so that inside the PPLN waveguide, its starting edge is matched with
the temporal phase carried by
the pump.  The PPLN output goes through a pair of prisms and a 750~nm short
pass filter to remove residual and frequency doubled pump light,
and the upconverted signal at 600~nm is analyzed via a second FP (10~GHz free spectral range and finesse $>$ 100).

\begin{figure}[t]
\centerline{\includegraphics[width=10 cm]{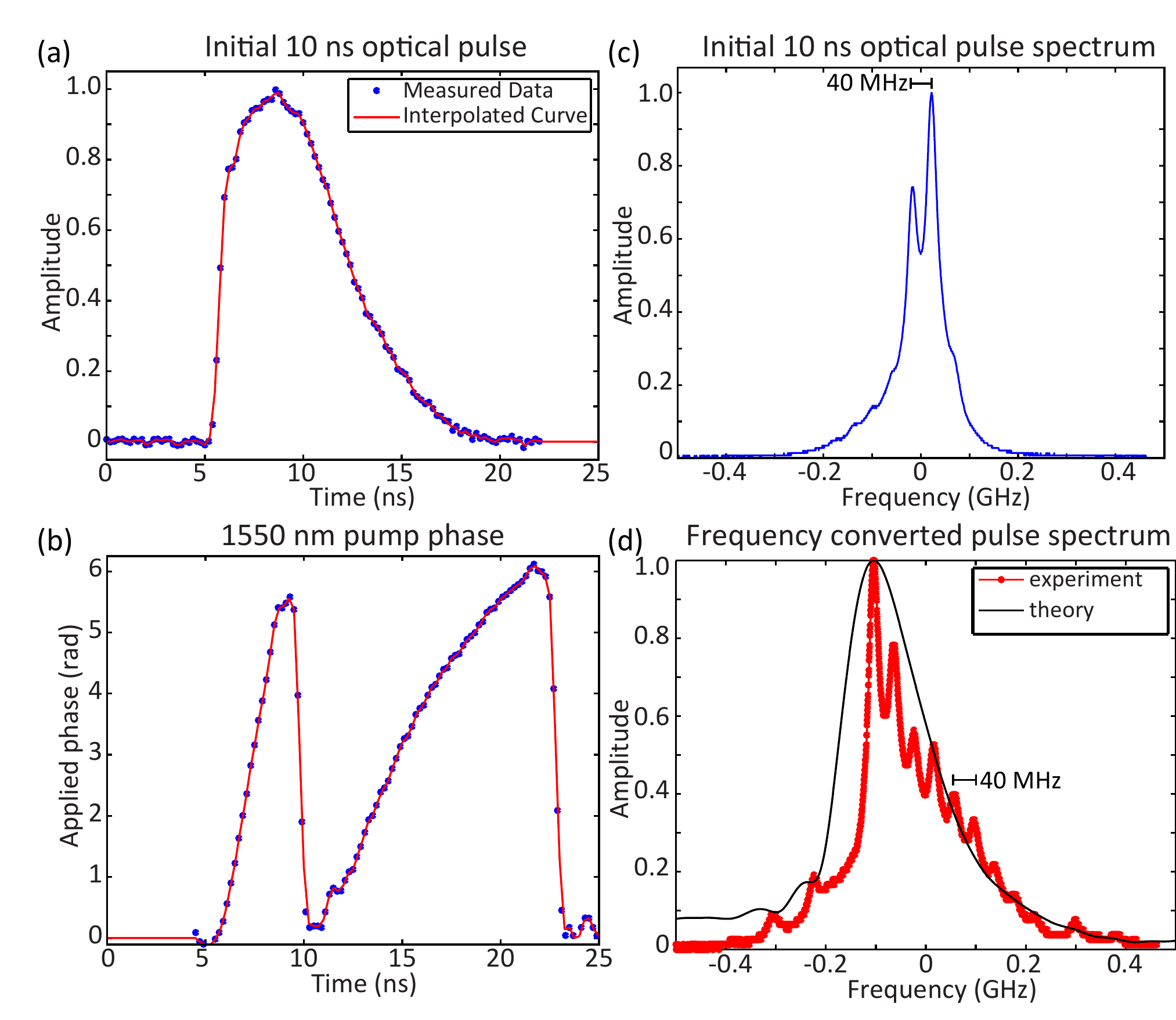}} \caption{(a) One period of a 40~MHz repetition rate, 980~nm input optical pulse train. (b) Applied phase to convert the spectrum of the 10 ns mono-exponential decay to the spectrum of a 1 ns Lorentzian. (c) FP-measured spectrum of the input 980 nm pulses. (d) The frequency converted, spectrally shaped output
at 600 nm. The black line is the theoretically expected output (does not include the pulse repetition rate), while the red curve is the experimental output measured by the FP analyzer.}\label{fig:Fig2}
\vspace{-0.1in}
\end{figure}

Figure~\ref{fig:Fig2} shows the data for spectral shaping of a 10 ns
mono-exponentially decaying pulse, similar to that
expected for the nitrogen vacancy center in diamond~\cite{ref:Kurtsiefer}) to the spectrum of
a 1 ns Lorentzian (a mono-exponential decay in frequency). The input optical pulse is
shown in Fig.~\ref{fig:Fig2}(a), while the temporal phase function needed for
shaping is shown in Fig.~\ref{fig:Fig2}(b).  As this phase function exceeds $2\pi$
half way through the pulse, and the AWG/EOM can supply a
maximum of $2\pi$ phase, the phase is wrapped around near the $2\pi$ point.
Figure~\ref{fig:Fig2}(c) shows the measured spectrum for the initial 980~nm
pulse, a Lorentzian in which the two superimposed peaks result from the pulse repetition rate
(40~MHz), and subsidiary shoulders are due to imperfections in the
FP response. Figure~\ref{fig:Fig2}(d) shows the measured
spectrum of the frequency converted signal. The black line is the theoretically calculated
spectrum for the 10~ns to 1~ns spectral broadening and shaping, given the
input 980~nm pulse and applied temporal phase. The red points are
the experimental data measured via the second FP analyzer. The
peaks in the frequency converted 600~nm spectrum are due to the 40~MHz pulse
repetition rate. The envelope of these peaks matches the
theoretical curve (which does not include the repetition rate) well.

This agreement can be quantified by an overlap integral ($I$) between
the data ($|\Omega_{\text{exp}}(\omega)|$) and theory ($|\Omega_{\text{th}}(\omega)|$),
defined as:

\begin{equation}
I=\frac{(\int|\Omega_{\text{exp}}(\omega)||\Omega_{\text{th}}(\omega)|d\omega)^2}{\int|\Omega_{\text{exp}}(\omega)|^2d\omega\int|\Omega_{\text{th}}(\omega)|^2d\omega}
\end{equation}

\noindent Including the pulse repetition rate and FP response, we find $I=0.88\pm~0.02$, indicating some success in
this spectral shaping approach~\cite{ref:overlap_note}. Deviation from $I$=1 is likely due to a number of factors, including the limited sampling rate of our AWG and the corresponding distortions it causes in the applied phase, particularly in the 'wrapping around' points (when the phase reaches 2$\pi$).

\begin{figure}[t]
\centerline{\includegraphics[width=7.5 cm]{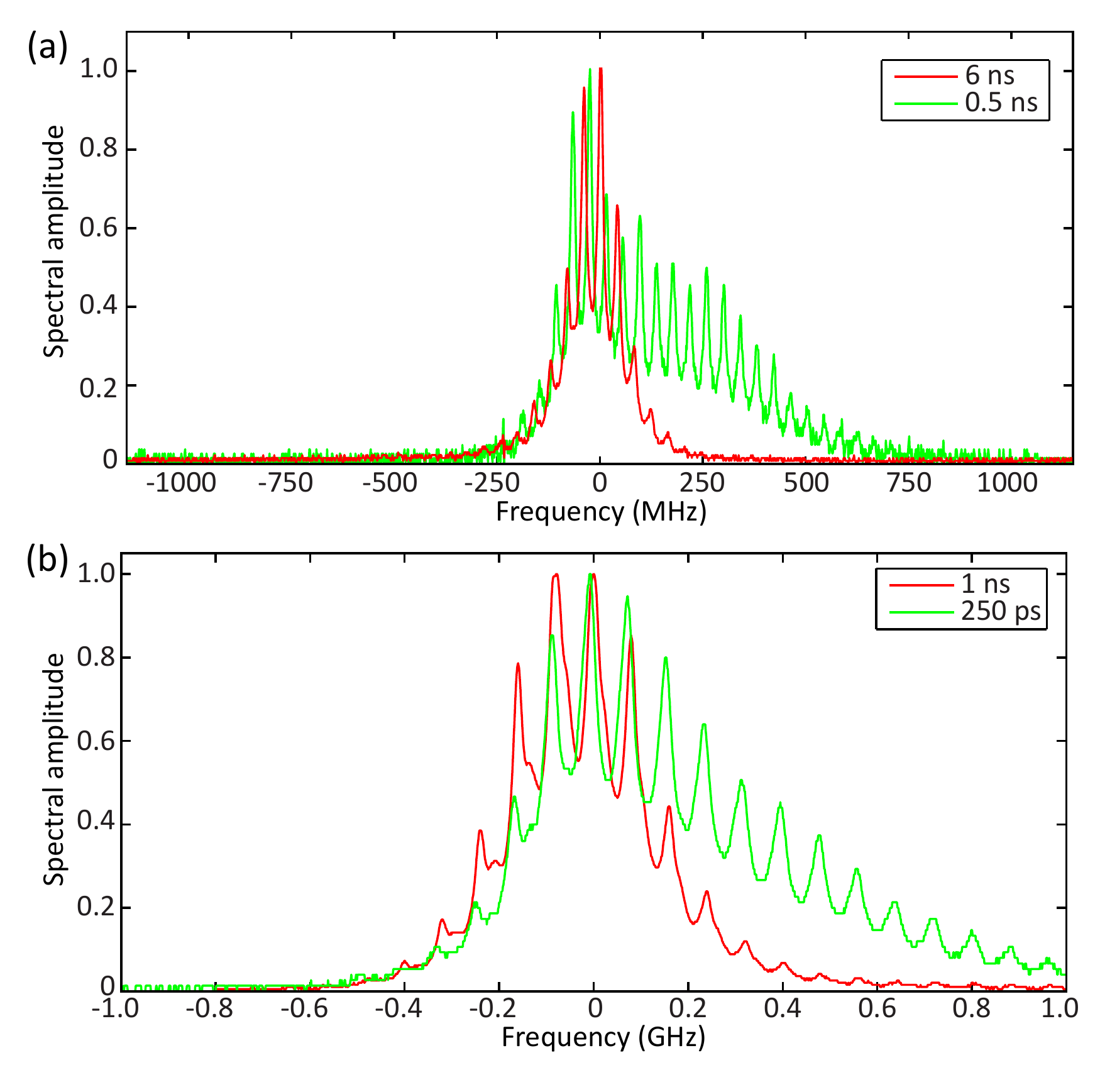}} \vspace{-0.2in} \caption{(a) Broadening the spectrum of a 6~ns exponentially
decaying pulse to the spectrum of a 0.5~ns Lorentzian. (b) Same as (a), but
from 1 ns to 250 ps.} \vspace{-0.1in} \label{fig:Fig3}
\end{figure}

One feature of this spectral shaping approach is the ability to
configure it to convert a variety of different input spectra to desired
output spectra. In Fig.~\ref{fig:Fig3} we demonstrate the spectral broadening and shaping of
input pulse durations that are consistent with those measured for other
single quantum emitters. In Fig.~\ref{fig:Fig3}(a), we show the conversion
of a 6~ns decaying pulse, similar to that of organic dye
molecules used as single photon sources~\cite{ref:Lettow_Sandoghdar_OE}, to
the spectrum of a 0.5~ns Lorentzian pulse. In Fig. 3(b), we show the
conversion of the spectrum of a 1~ns pulse, similar to that from quantum
dot single photon sources~\cite{ref:Michler_book_2009},
to the spectrum of a 250 ps pulse, which is at the lower limit of output
pulse duration that can be achieved given the sampling rate of our AWG.

%There are a number of potential applications of
%this spectral broadening and shaping, if combined with spectral phase
%correction to achieve temporal compression, as described above.  For example,
%one can envision using it as a means to generate indistinguishable photons
%from disparate quantum emitters. It might also be useful as a route to
%increasing the speed and brightness of single photon sources produced by
%system like neutral atoms, e.g., if full waveform shaping is combined with
%multiplexing of several sources.  Such waveform-shaped photons would also be
%compatible with broadband quantum memories.

\section{Towards temporal waveform shaping}

To move from spectral shaping to temporal shaping, residual
spectral phase must be removed from the generated pulse, and can be done by a number
of techniques~\cite{ref:Weiner_RSI}. Here, we provide numerical calculations
of the full spectro-temporal waveform shaping protocol.

\begin{figure}[ht]
\centerline{\includegraphics[width=\linewidth]{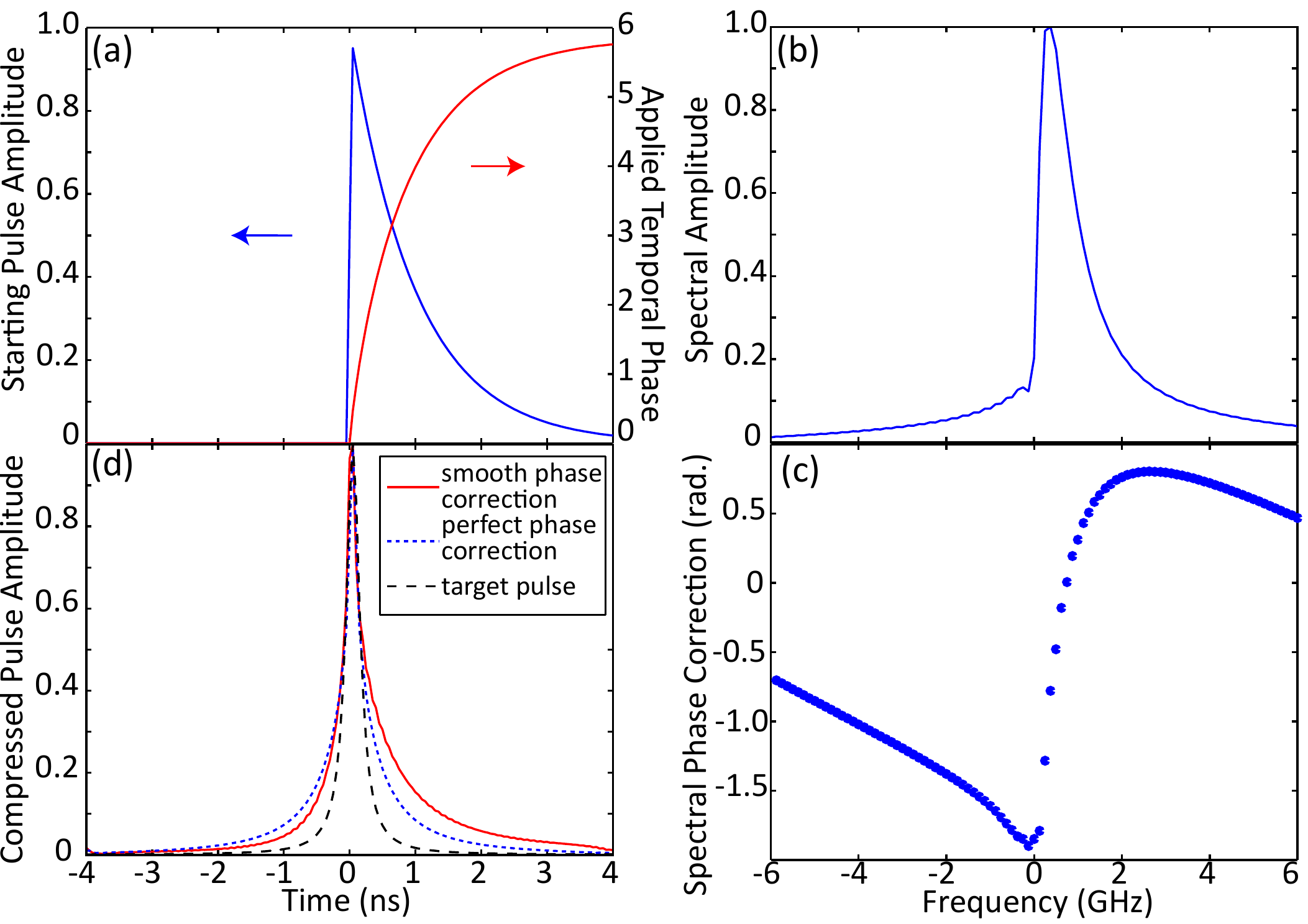}} \vspace{-0.1in} \caption{Towards full spectro-temporal shaping (calculations).  Clockwise from the top left: (a) Initial 1~ns mono-exponetially decaying pulse (blue) and applied temporal phase (red). (b) Spectrum after temporal phase application. (c) Smoothed spectral phase correction for temporal shaping. (d) Temporally shaped pulse (red), along with result for perfect spectral phase correction (blue dotted line) and the target 250~ps Lorentzian pulse (black dashed line). \vspace{-0.2in}}\label{fig:Fig4}
\end{figure}

We target compression of a 1~ns mono-exponential decay (Fig.~\ref{fig:Fig4}(a), blue
line) to a 250~ps (full-width at half-maximum) temporal Lorentzian (Fig.~\ref{fig:Fig4}(d), black
dashed line). First, we calculate the appropriate temporal phase (Fig.~\ref{fig:Fig4}(a), red line)
to create the spectrum of the 250~ps Lorentzian (Fig.~\ref{fig:Fig4}(b)). Achieving flat phase across
this spectrum requires a spectral phase correction function that is highly oscillatory.  A smoothed function that
should be easier to implement in practice is shown in Fig.~\ref{fig:Fig4}(c).  Figure~\ref{fig:Fig4}(d) shows
the resulting temporally compressed waveform (red line), along with the waveform that would be generated if
perfect spectral phase correction is achieved (blue dotted line).  Both of these curves are distorted from the
target 250~ps Lorentzian (black dashed line), indicating that the spectral
phase compensation and temporal phase application steps are not ideal.  This
is due to various approximations used, such as the method of stationary phase and our choice of smoothly varying functions.
The resulting compressed waveform has a full-width at half-maximum of $\approx$~310~ps, indicating that despite the distortion,
a significant temporal re-shaping and compression is feasible.

In summary, we have implemented the first stage of a proposal for
quantum waveform shaping~\cite{ref:Kielpinski_PRL_2011}, where nonlinear mixing of an input mono-exponentially decaying pulse with a phase-modulated
pump simultaneously translates the wavelength of the pulse and spectrally broadens and shapes it to match a desired
output spectrum. This approach is compatible with single photons and adds to the toolkit of resources
being developed for applications in photonic quantum information science.

I.A, S.A., and L.S. acknowledge support under the Cooperative Research Agreement
between the University of Maryland and NIST-CNST, Award
70NANB10H193.

%\bibliographystyle{osajnl}
%\bibliography{KS_bib_2014_1_31}

\begin{thebibliography}{10}
\bibitem{ref:spectral_broadening_note}
Conversion efficiency is taken as the ratio of frequency converted photon flux to input photon flux, taking into account the quantum efficiency of the single
photon counter at the two wavelengths. The uncertainty is the one standard deviation value and is due to fluctuations in the detected count rate.
\bibitem{ref:overlap_note}
The uncertainty in the overlap integral is due to possible temporal misalignment between
the 1550~nm and 980~nm signals in the PPLN waveguide.  We estimate this to be $\pm$250~ps (one standard deviation value).
\end{thebibliography}

\end{document}